\documentclass[letterpaper]{article} 
\usepackage{aaai2026}  
\nocopyright
\usepackage{times}  
\usepackage{helvet}  
\usepackage{courier}  
\usepackage[hyphens]{url}  
\usepackage{graphicx} 
\usepackage{natbib}  
\usepackage{caption} 
\usepackage{algorithm}
\usepackage{algorithmic}

\usepackage{algorithm}
\usepackage{algorithmic}
\usepackage{amsmath}
\usepackage{amsthm}
\usepackage{amssymb}
\usepackage{booktabs} 
\usepackage{array}

\newtheorem{proposition}{Proposition}

\usepackage{newfloat}
\usepackage{listings}
\DeclareCaptionStyle{ruled}{labelfont=normalfont,labelsep=colon,strut=off} 
\floatstyle{ruled}
\newfloat{listing}{tb}{lst}{}
\floatname{listing}{Listing}
\title{VIL2C: Value-of-Information Aware Low-Latency Communication for Multi-Agent Reinforcement Learning}
\author {
   Qian Zhang\textsuperscript{\rm 1},
   Zhuo Sun\textsuperscript{\rm 1}\thanks{Corresponding author: zsun@nwpu.edu.cn},
    Yao Zhang\textsuperscript{\rm 1},
    Zhiwen Yu\textsuperscript{\rm 1, 2},
    Bin Guo\textsuperscript{\rm 1},
    Jun Zhang\textsuperscript{\rm 3}
}
\affiliations {
    \textsuperscript{\rm 1}Northwestern Polytechnical University\\
    \textsuperscript{\rm 2} Harbin Engineering University\\
    \textsuperscript{\rm 3}The Hong Kong University of Science and Technology\\ 
    qianz5152@mail.nwpu.edu.cn, zsun@nwpu.edu.cn, yaozh.g@nwpu.edu.cn, zhiwenyu@nwpu.edu.cn, guob@nwpu.edu.cn, eejzhang@ust.hk
}

\usepackage{bibentry}

\begin{document}

\maketitle

\begin{abstract}
Inter-agent communication serves as an effective mechanism for enhancing performance in collaborative multi-agent reinforcement learning(MARL) systems. However, the inherent communication latency in practical systems induces both action decision delays and outdated information sharing, impeding MARL performance gains, particularly in time-critical applications like autonomous driving. In this work, we propose a \textbf{Value-of-Information aware Low-latency Communication(VIL2C)} scheme that proactively adjusts the latency distribution to mitigate its effects in MARL systems. Specifically, we define a Value of Information (VOI) metric to quantify the importance of delayed message transmission based on each delayed message's importance. Moreover, we propose a progressive message reception mechanism to adaptively adjust the reception duration based on received messages. We derive the optimized VoI aware resource allocation and theoretically prove the performance advantage of the proposed VIL2C scheme. Extensive experiments demonstrate that VIL2C outperforms existing approaches under various communication conditions. These gains are attributed to the low-latency transmission of high-VoI messages via resource allocation and the elimination of unnecessary waiting periods via adaptive reception duration.


\end{abstract}

\section{Introduction}
Multi-Agent Reinforcement Learning (MARL) has shown impressive potential in solving complex multi-agent decision-making problems across real-world applications, such as autonomous driving \cite{Palanisamy:autonomousvehicles}, traffic control \cite{Wei:trafficcontrol}, and robot control \cite{Gu:robot}. While pioneering studies devote considerable efforts to develop algorithms under the centralized training and decentralized execution (CTDE) paradigm \cite{yu:mappo}, MARL still suffers from the challenges caused by the non-stationary \cite{oroojlooy:pons} and partially observable environments \cite{foerster:rial}.

\begin{figure} [t]
\centering
\includegraphics[width=0.35\textwidth]{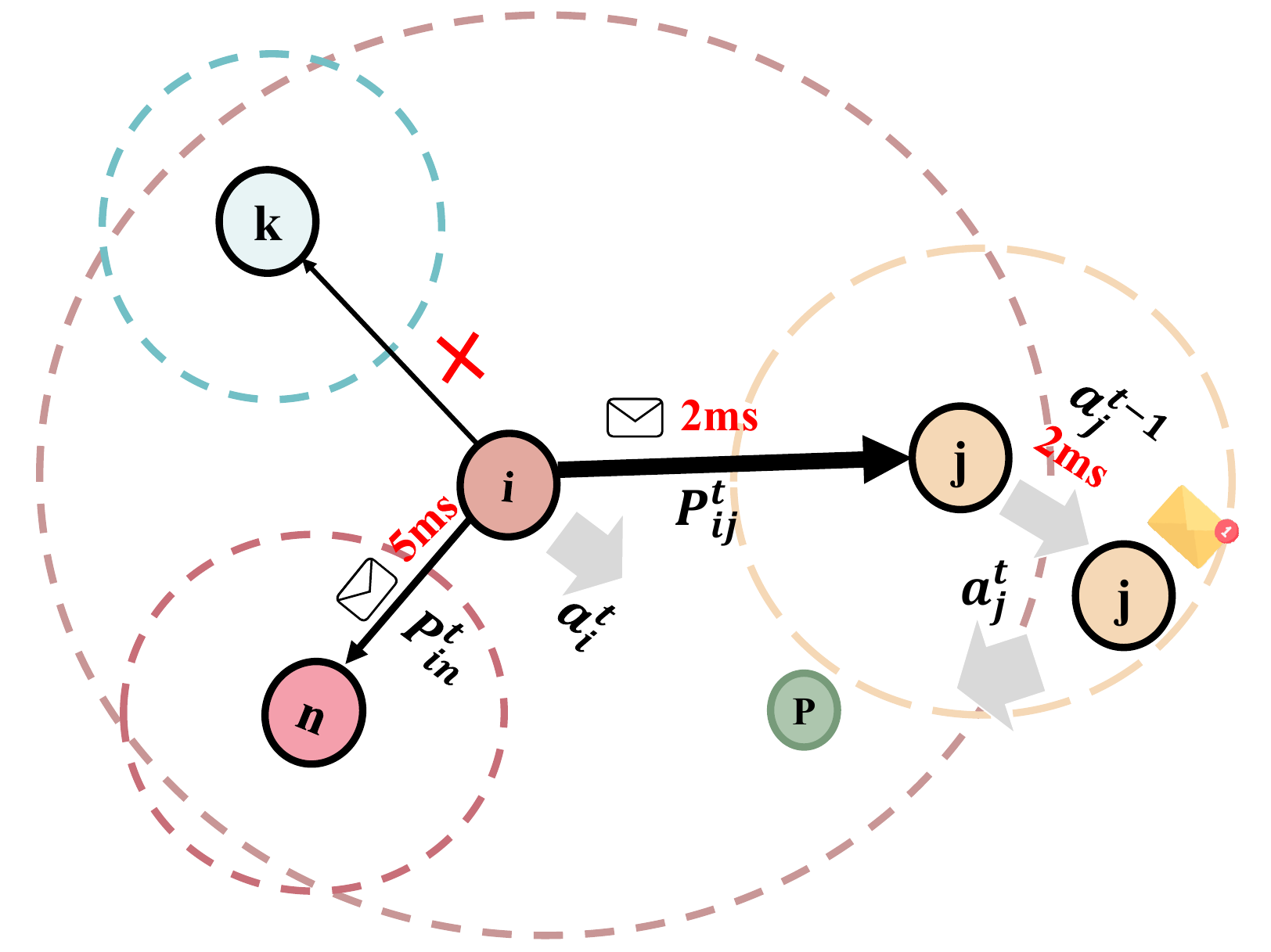}
\caption{The toy example illustration of VIL2C. Agent $i$ transmits messages to agents $j$ and $n$ to collaboratively capture target $P$. Based on the message importance relative to recipients $j$ and $n$, agent $i$ allocates transmission resources (represented by arrow thickness) to proactively adjust latency. Agent $j$ terminates reception (represented by red dot), once receiving sufficient messages. Black arrows represent communication links and gray arrows represent movement.}
\label{fig1:env}
\end{figure}
To address these challenges, inter-agent communication is incorporated into MARL~\cite{zhu:comsurvey}. Early works \cite{Zhang:broadcast} adopt the broadcast mechanism, allowing agents to share messages with each other, which incurs substantial communication overheads. Resource constraints in communication systems hinder the deployment of broadcast mechanisms in real-world multi-agent systems. Thus, several studies \cite{Hu:etcnet,sun:t2m2c,guo:tem,Li:cacom,wang:imac} explore effective communication protocols, where agents dynamically determine the communication time, partners, and messages based on demands. Although these protocols effectively reduce communication overheads, they ignore the inevitable and dynamic communication latency in real-world systems.  

Communication latency may impair multi-agent cooperation through causing action decision delays \cite{chen:2020,chen:2021} and outdated information sharing. Particularly, the latency of decision-critical messages dominates this impact. Previous work \cite{yuan:dacom} avoids unbounded waiting by dynamically adjusting the waiting time based on real-time network conditions. CoDe \cite{song:code} alleviates misguidance of outdated messages by inferring the long-term intent and prioritizing the most relevant recent messages. While these approaches mitigate the latency impact on cooperation performance via well-designed reception schemes, they rely on the channel-dependent inherent latency distribution but overlook the potentials of proactive latency distribution adjustment. 
Proactive and precise adjustment of latency distribution can be achieved by effective resource allocation in wireless communication systems \cite{Sun19}.
Unlike conventional wireless communication systems that focus on maximizing sum rates or minimizing packet-level latency \cite{She17,Liu23,Meng22}, the communication design in MARL requires considering the heterogeneous importance of delayed messages on recipients' decisions to improve cooperation performance. 
Thus, a novel communication mechanism is needed to adapt latency according to message importance under latency awareness.
  
In this work, we propose a Value-of-Information aware Low-Latency multi-agent Communication (VIL2C) scheme. Firstly, we define Value of Information (VoI) in MARL to quantify how delayed messages influence recipient agents' action decisions, accounting for both their semantic importance and associated communication latency. Based on the defined VoI, we jointly design a message transmission scheme and a progressive reception strategy. The message transmission scheme dynamically optimizes both bandwidth and power allocation by jointly considering the heterogeneous VoI of different messages and time-varying wireless channels. Such resource allocation can reduce communication latency of messages that critically influence the recipient's action decision, thereby mitigating the impact on multi-agent cooperation performance. Meanwhile, the progressive message reception strategy adaptively determines the recipient's waiting time of communication, ceasing reception once sufficient information for reliable action decision is obtained. This strategy aims to minimize the waiting duration while making a reliable action decision. A toy example is shown in Fig.~\ref{fig1:env}, where agent $i$ prioritizes resource allocation for transmitting messages to agent $j$ and agent $j$ terminates its reception process upon the receipt of sufficient messages. Moreover, we provide theoretical analysis that establishes the VIL2C's performance guarantee in enhancing multi-agent cooperation. We evaluate the proposed VIL2C scheme on three typical multi-agent environments: Predator-Prey (PP), Cooperative Navigation (CN) \cite{lowe:mpe}, and Starcraft Multi-Agent Challenge (SMACv2) \cite{ellis:smacv2}. Extensive experiment results demonstrate that VIL2C outperforms all considered baselines. We also apply VIL2C and baselines under different resource budgets and channel conditions. The experiments show that VIL2C exhibits strong robustness against various communication conditions. The contributions of this work are summarized as follows:

\begin{itemize}

\item To the best of our knowledge, this work is the first attempt to proactively adjust communication latency distribution in MARL by VoI aware resource allocation.




\item We propose a VoI aware low-latency communication system in MARL, named VIL2C, which innovatively introduces VoI into the designs of message transmission and reception to improve cooperation performance with considering communication latency.

\item We derive the optimized VoI aware resource allocation in VIL2C. We also provide theoretical analysis that establishes the VIL2C’s performance guarantee in enhancing multi-agent cooperation performance.

\item We test the proposed VIL2C scheme in three typical multi-agent environments. VIL2C consistently demonstrates superior performance compared with baselines across various resource budgets and channel conditions.



\end{itemize}

\section{Related Works} 
Learning communication protocols in collaborative MARL has drawn significant attention, as strategic information exchange enhances agents' environmental awareness and cooperative performance. RIAL and DIAL \cite{foerster:rial} introduced end-to-end communication strategies, followed by broadcast mechanisms \cite{sukhbaatar:commnet,peng:bicnet,das:tarmac}. Although simple to implement, broadcast mechanisms incurs high overheads, motivating the development of more effective protocols.

Effective multi-agent communication protocols primarily address three key aspects: when to communicate, whom to communicate with, and what information to transmit. For timing, Gated-ACML \cite{mao:gateacml} and IC3Net \cite{singh:ic3net} employed gating mechanisms, while ETCNet \cite{Hu:etcnet} used event-triggered networks to decide whether and whom to engage. For recipient selection, I2C \cite{ding:i2c} and ToM2C \cite{wang:tom2c} leveraged causal inference. TEM \cite{guo:tem} defined Kullback-Leibler (KL) divergence based importance metric to select transmissions. For content, T2MAC \cite{sun:t2m2c} supported selective engagement and evidence-driven integration. IMAC \cite{wang:imac} applied information bottleneck for compact messages. The studies \cite{wang:ndq,yuan:maic,Li:cacom} generated personalized messages tailored to recipients' needs. While the protocols reduce communication overhead, they neglect latency's impact.

Communication latency is unavoidable in practical multi-agent systems. DA-MG \cite{chen:2020} modeled delay aware Markov games to assess action delays' impact on decision quality. DACOM \cite{yuan:dacom} integrated latency into MARL with adaptive waiting time scheduling for dynamic networks. CoDe \cite{song:code} addressed latency across decision intervals via dual alignment for asynchronous message integration. These methods rely on channel-dependent latency distributions, whereas our work explores proactive adjustment through joint message transmission and reception design.

\begin{figure*}[t] 
    \centering
    \includegraphics[width=\textwidth]{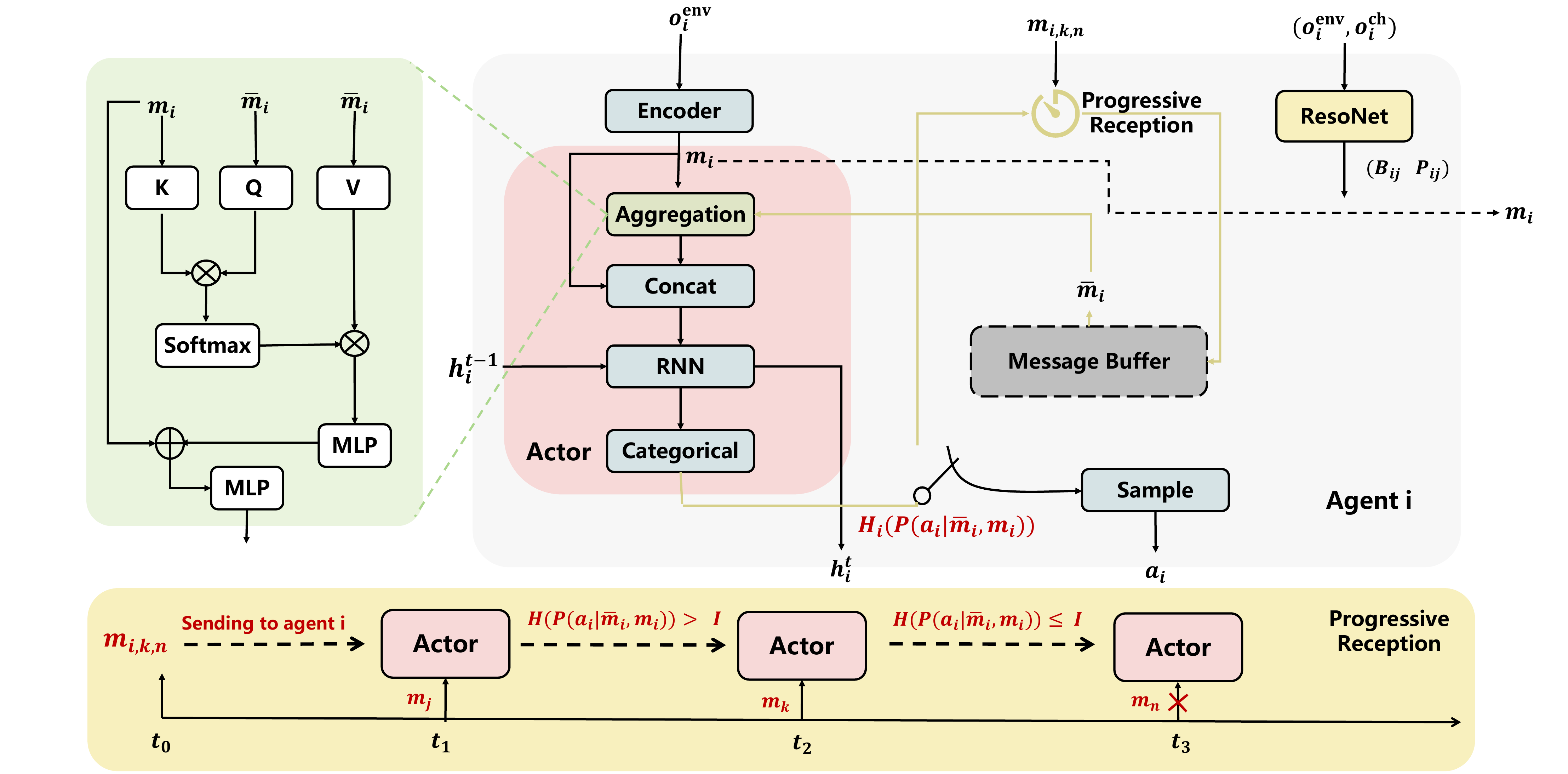}
    \caption{The framework of VIL2C. Each agent consists of six components: 1) an Encoder that generates the message from the local observation, 2) a ResoNet that performs the online resource allocation, 3) a Message Buffer that stores received messages, 4) an Actor that utilizes the own message and received messages to obtain an action probability distribution, 5) a Progressive Reception module that determines whether to stop receiving messages, illustrated by the yellow box, and the corresponding process is represented by yellow lines, 6) a Sample module to select actions from the action probability distribution.}
    \label{fig2:framework}
\end{figure*}
\section{Background}
\subsection{Delay aware Decentralized Partially Observable Markov Decision Process}
The Delay aware Decentralized Partially Observable Markov Decision Process (Da-Dec-POMDP) extends the standard Dec-POMDP framework to explicitly model communication latency in multi-agent systems. Formally, the Da-Dec-POMDP is defined as $\mathcal{G} = \langle \mathcal{S}, \mathcal{A}, \mathcal{O}, P, R, \Omega, N, \gamma, \mathcal{T} \rangle$. Here, $N$ is the number of agents in the system, $\mathcal{S}$ is the global state space, $\mathcal{A}$ is the joint action space of all agents, $\mathcal{O}$ is the partial observation space for agents. $P$ represents the transition function characterizing the effect of joint action on states, $R$ represents the reward function used to assess the quality of action decisions in current states, and $\Omega$ represents the observation function mapping states to observations. Meanwhile, $\gamma$ is the discount factor and $\mathcal{T}$ is the communication latency distribution. At timestep $t$, agent $i$, $i\in [1,...,N]$, obtains the local observation $o_i^t = (o_i^{\text{env}}, o_i^{\text{ch}})$ that captures both environment and channel information. It then transmits its observation to the selected agent $j$ with a random latency $\tau_{i,j} \sim \mathcal{T}_{i,j}$. Based on the local observation and received messages, agent $i$ decides its action $a^t_i$ according to the policy $\pi_i$. The joint action of all agents $\mathbf{a}^t$ leads to state transition and generates the global reward $r^t = R(s^t, \mathbf{a}^t)$. The objective is to learn the joint policy $\boldsymbol{\pi}$ that maximizes the expected discounted return $J(\boldsymbol{\pi}) = \mathbb{E}_{\boldsymbol{\pi}} \left[ \sum_{t=0}^{\infty} \gamma^t r^t \right]$ under the stochastic dynamics induced by policies, transitions, and communication latency.

\subsection{Communication Channel Model}
In real-world multi-agent cooperation scenarios, the agents exchange messages via a wireless communication network. In wireless networks, the point-to-point communication latency between two agents is mainly determined by the transmission rate for a given message size. According to \cite{gold:rij}, the transmission rate from agent $i$ to agent $j$ is given by
\begin{equation}
R_{i,j} = B_{i,j} \log_2 \left( 1 + \frac{P_{i,j} }{10^{\frac{PL_{i,j}}{10}} B_{i,j}N_0} \right),
\label{eq:capacity}
\end{equation}
where \( B_{i,j} \) and \( P_{i,j} \) refer to the bandwidth (Hz) and transmit power (W) allocated by agent $i$ for transmission to agent $j$, respectively, and \( N_0 \) is the power spectral density of additive white Gaussian noise (AWGN). Here, \( PL_{i,j} \)(in dB) refers to the distance-dependent path loss used to measure the signal propagation attenuation in wireless channels, modeled as
\begin{equation}
PL_{i,j} = \eta \log_{10}d_{i,j} + c,
\end{equation}
where \( d_{i,j} \) is the distance between agent \( i \) and agent \( j \), and \( c \) is the offset term representing other fixed effects. \( \eta \) represents the path loss exponent determined by the communication environment, such as $\eta=2$ for the free space and $\eta=3.67$ for the urban area \cite{cho:mimo}. It can be seen that the available transmission rate depends on the transmit agent's bandwidth and transmit power allocation under the given channel condition. 

\section{Methodology}
In this section, we present the detailed design of VIL2C. As the communication module, VIL2C works with the action decision module in multi-agent proximal policy optimization (MAPPO), which is a CTDE method. Since it is parallel to the action decision module, VIL2C is compatible to any CTDE method.

\subsection{Value-of-Information aware Low-latency Communication}

In this part, we first define VoI to measure the importance of delayed messages on the recipient agent's action decision. Based on VoI, we propose the VIL2C module, which consists of a message transmission scheme and a progressive message reception strategy, as shown in Fig.\ref{fig2:framework}. We also derive a lower bound of cooperation performance in the presence of communication latency and theoretically prove the performance improvement of VoI aware delay distribution adjustment obtained by the proposed VIL2C.


\noindent \textbf{Value of Information.} 
In cooperative MARL scenarios, the message importance relative to a recipient agent is evaluated by the influence of its semantic information on this agent's action decision. A message is considered highly important when it significantly influences the recipient agent’s action decision. Otherwise, it is deemed low importance. On the other hand, receiving the message incurs latency as a cost. We define VoI as the message importance per latency cost, given by

\begin{equation}
\label{eq:voi}
VoI_{i,j} = \frac{\xi_{i,j}}{\tau_{i,j}},
\end{equation}
where $\xi_{i,j}$ represents the influence of agent $i$'s message \( m_i \) on recipient agent $j$'s action decision and $\tau_{i,j}$ is communication latency of message \( m_i \) to recipient agent $j$. To quantify the influence of message \( m_i \) on recipient agent $j$'s action decision, we utilize KL divergence to measure the discrepancy in the action probability distribution of recipient agent \( j \) before and after receiving message \( m_i \), expressed as
\begin{equation}
\xi_{i,j} = D_{KL}\big(P(a_j \mid o_j, m_i) \,||\, P(a_j \mid o_j)\big),
\label{kl}
\end{equation}
where \( P(a_j \mid o_j) \) and \( P(a_j \mid o_j, m_i) \) denote the action probability distribution of agent \( j \) before and after receiving message \( m_i \), respectively. The communication latency of message \( m_i \) to agent \( j \) is
\begin{equation}
\tau_{i,j} = \frac{L_i}{R_{i,j}},
\label{latency}
\end{equation}
where \( L_i \) is the size of message \( m_i \) in bits and \( R_{i,j} \) is the available transmission rate from agent \( i \) to \( j \). From Eqs. \eqref{eq:capacity} and \eqref{latency}, it is observed that bandwidth and transmit power allocation can affect available transmission rates, thereby reshaping the inter-agent communication latency distribution.

\noindent \textbf{VoI aware Transmission.}
To reshape the inter-agent communication latency distribution, i.e., reducing communication latency of important messages, we propose the VoI aware message transmission scheme. In this scheme, the transmit agent optimizes its bandwidth and power allocation to maximize the total VoI of its transmissions. This resource allocation problem is formulated as 
\begin{equation}\label{reso_all}
\arg \max_{\{B_{i,j},P_{i,j}\}} \sum_{j=1}^{N_i} VoI_{i,j}(B_{i,j},P_{i,j}),
\end{equation}
\[
\begin{aligned}
\text{s.t.} \quad & \sum_{j=1}^{N_i} P_{i,j} \leq P_{\text{budget}}, 
& \sum_{j=1}^{N_i} B_{i,j} \leq B_{\text{budget}},
\end{aligned}
\]
where $N_i$ is the number of agent $i$'s recipients and $VoI_{i,j}(B_{i,j},P_{i,j})$ is obtained by incorporating Eqs. \eqref{eq:capacity}, \eqref{kl}, and \eqref{latency} into \eqref{eq:voi}. Here, $B_{i,j}$ and $P_{i,j}$ are the bandwidth and transmit power allocated by agent \(i\) for the transmission to agent \(j\), respectively. \(B_{budget}\) and \(P_{budget}\) are the transmit agent's available bandwidth and power budget, respectively. We first consider that each agent can obtain the KL divergence based importance of its message on recipients' action decisions before transmissions. In this case, the optimization problem is solved by using Karush-Kuhn-Tucker (KKT) \cite{Kuhn2014} conditions in Proposition \ref{prop:value-optimization}. 

\begin{proposition}
\label{prop:value-optimization}
Consider the importance $\xi_{i,j}$ of agent $i$'s message on recipient $j$. The optimized bandwidth $B_{i,j}^*$ and transmit power $P_{i,j}^*$ allocated by agent $i$ to agent $j$ satisfy
\begin{subequations}
\begin{align}
\frac{\xi_{i,j}}{L_{i}}\left( \log_2(1 + \gamma_{i,j}) - \frac{\gamma_{i,j}}{1 + \gamma_{i,j}} \right) = \lambda, \label{eq:kkt_p_full}  \\
\frac{\xi_{i,j} \gamma_{i,j} B_{i,j}^*}{L_{i}P_{i,j}^*(1 + \gamma_{i,j})} = \mu. \label{eq:kkt_b_full}    
\end{align}
\end{subequations}
respectively, and $B_{i,j}^*$ and $P_{i,j}^*$ are proportional to $\xi_{i,j}$. Here, $\gamma_{i,j} = \frac{P_{i,j}^*}{10^{\frac{PL_{i,j}}{10}}B_{i,j}^* N_0 }$, $\tau_{i,j}$ is communication latency from agent $i$ to $j$, defined as Eq. \eqref{latency}. $\mu$ and $\lambda$ are Lagrange multipliers for bandwidth and power constraints, respectively.
\end{proposition}
It is concluded from Proposition \ref{prop:value-optimization} that it preferentially allocates bandwidth and power to critical messages, thereby reducing their communication latency.

In practice, it is difficult for agents to obtain the importance of their messages before transmission. To address this, we design a neural network within VIL2C to optimize resource allocation, called \textbf{ResoNet}. By exploiting the CTDE method, ResoNet is first trained to maximize the total VoI based on global information and message importance in a centralized way. Then, ResoNet uses local state information, including local observation and wireless channels to other agents, as input and decides the optimized bandwidth and power allocation in a decentralized way, as shown in Fig. \ref{fig2:framework}. The detailed network design will be presented next.

\noindent  \textbf{Progressive Reception.}
The agent waits for receiving messages from multiple agents. Due to the existence of communication latency, messages arrive asynchronously. The reception of more messages means longer waiting duration, which may cause unaffordable action delays and task failure. Conversely, if the waiting duration is too short, it may miss many messages and degrade multi-agent cooperation performance. Thus, we propose a progressive reception strategy to minimize the waiting duration while ensuring a reliable action decision. Specifically, the recipient agent immediately processes this message and obtains the corresponding action probability distribution, once one message is received. When the uncertainty of its action probability distribution is smaller than a predefined threshold, the agent terminates its reception and takes an action. Otherwise, the agent continues waiting for more messages. We adopt the entropy to quantify the uncertainty of an action probability distribution. The termination condition of message reception is

\begin{equation}
H(P(a \mid \bar{m}_i,m_i) ) \leq I,
\end{equation}
where $H(\cdot)$ is the entropy function and \(P(a \mid \bar{m}_i, m_i)\) is
agent $i$’s action probability distribution conditioned on messages of the buffer $\bar{m}_i$ and its own message $m_i$. Here, \( I \) is a predefined entropy threshold, indicating the maximum acceptable uncertainty of action probability distribution to obtain a satisfactory action decision.
It is obvious that a larger entropy threshold leads to the reception of more messages and longer waiting, and vice versa. Then, there exists an optimal entropy threshold to balance two effects and to achieve the maximum reward. This optimal entropy threshold is obtained by the grid search from $0$ and $log_2{n}$, where $n$ is the dimension of action space in the considered task. In the proposed VIL2C, the message transmission scheme significantly reduces communication latency of important messages via preferential resource allocation. Thus, important messages are received earlier with a high probability and the termination condition can be satisfied in advance, thereby reducing the waiting duration. Note that, a maximum waiting time is set to prevent unbounded waiting. The effect of maximum waiting time will be evaluated in the experiments. We also note that when the recipient agent decides to terminate reception, it feeds back an ACK to notify other agents to halt their message transmission, thereby releasing communication resources~\cite{Ahmed:nack}. 

As illustrated in Fig.~\ref{fig2:framework}, agents transmit messages to agent $i$ at time $t_0$. Due to the presence of communication latency, agent $i$ receives the agent $j$'s message at time $t_1$. After processing the message, it checks whether the termination condition is satisfied. When the termination condition is not satisfied, agent \( i \) continues waiting for messages from other agents. At time $t_2$, agent $i$ receives the agent $k$'s message and processes it to check the termination condition. When it is satisfied, agent \( i \) terminates reception and takes an action.  

\noindent \textbf{Theoretical analysis.}
We use the cumulative reward as a performance metric to theoretically illustrate the advantage of VoI aware latency distribution adjustment in VIL2C. And, we first derive a lower bound for the cumulative reward under an average communication latency in Proposition \ref{prop:lower-bound}. 

\begin{proposition}
\label{prop:lower-bound}
Consider a cumulative reward $J_{\text{opt}}$ of MARL without considering communication latency. We have the following lower bound for the cumulative reward of communication latency aware MARL:
\begin{equation}
    J \geq J_{\text{opt}} - C \cdot \bar{\tau},
\end{equation}
where $\bar{\tau}$ is the mean of communication latency distribution and $C>0$ is a task-dependent constant representing the latency sensitivity. 
\end{proposition}

Then, we analyze the advantage of VoI aware latency distribution adjustment in VIL2C. In previous multi-agent communication works, where resources are equally allocated to transmission links and the channel-dependent inherent latency distribution is used, the cumulative reward is modeled as $J_r$. Compared to $J_r$, the cumulative reward improvement achieved by our latency distribution adjustment is derived in Proposition \ref{prop:value-prioritized-delay-bound}. 
\begin{proposition}
\label{prop:value-prioritized-delay-bound}
Given the average communication latency $\bar{\tau}$ and the reward $R$ at a time step, the cumulative reward difference between proposed distribution adjustment $J_v$ and the channel-dependent inherent latency distribution $J_r$ is
\begin{equation}
    J_v - J_r = \operatorname{Cov}\!\bigl(R,e^{-\lambda\bar{\tau}}\bigr),
\end{equation}
and $\operatorname{Cov}(R, e^{-\lambda \bar{\tau}})\!\!>\!0$. Here, $\operatorname{Cov}(\cdot,\cdot)$ is covariance function and $\lambda$ is a latency-dependent reward discount factor. 

\end{proposition}
It is observed from Proposition \ref{prop:value-prioritized-delay-bound} that the cumulative reward difference increases with the latency-dependent reward discount factor $\lambda$. It implies that the cumulative reward improvement achieved by our distribution adjustment is more obvious in environments with high latency sensitivity. 

\subsection{Network Design}
The schematics of VIL2C is illustrated in Fig.~\ref{fig2:framework}. At each time step, agent $i$ obtains its local environment observation \( o_i^{\text{env}} \) and generates message $m_i$ via the Encoder. At the same time, agent $i$ uses the ResoNet, which inputs local environment observation \( o_i^{\text{env}} \) and wireless channel condition \( o_i^{\text{ch}} \), to obtain the optimal bandwidth and power allocation $(B_{i,j}^*,P_{i,j}^*)$. 

As a receiver, agent $i$ stores incoming messages in buffer $\bar{m}_i$ and uses an attention-based aggregator to fuse them with its own message $m_i$, yielding aggregated message $m_i^a$. Specifically, fully connected layers generate keys $k$ and values $v$ from $\bar{m}_i$, and query $q$ from $m_i$. Attention weights $\alpha$ are computed from $k$ and $q$ to capture correlations, then multiplied with $v$ to produce $m_i^a$ via a final fully connected layer. This is concatenated with $m_i$ to form $m_i^c$. An RNN processes $m_i^c$ and previous hidden state $h_i^{t-1}$ for actions, outputting to a Categorical module for action probability distribution. If the action distribution's entropy falls below the threshold, reception terminates and an action is sampled. Otherwise, the reception continues for more messages.


\subsection{Overall Training Objective}
Building upon the MAPPO framework \cite{yu:mappo}, our model is trained by optimizing two separate loss functions in an alternating fashion. Specifically, the loss of the action network in MAPPO is defined as
\begin{equation}
 \!\!\!\mathcal{L}_a(\theta)=\! \sum_{i=0}^N \left[ \min \left( r^i A^i, \text{clip}(r^i, 1\!-\!\epsilon, 1\!+\!\epsilon) A^i \right) \right],
\end{equation}
where \( r^i = \frac{\pi(a^i | o^i)}{\pi_{{\text{old}}}(a^i | o^i)} \) and \( A^i\) is the advantage function. 
In each training iteration, after updating the joint policy network, each agent's ResoNet is also updated. The training objective for ResoNet is to maximize the total VoI in Eq.~\eqref{reso_all}. To achieve this, we define the loss function for agent $i$ as the negative of the total VoI:
$\mathcal{L}^i_v(\theta) = -\sum_{j=1}^{N_i} VoI_{i,j}$.
The loss of critic network is same as MAPPO.

\begin{figure}[t]
    \centering
    \includegraphics[width=0.95\linewidth]{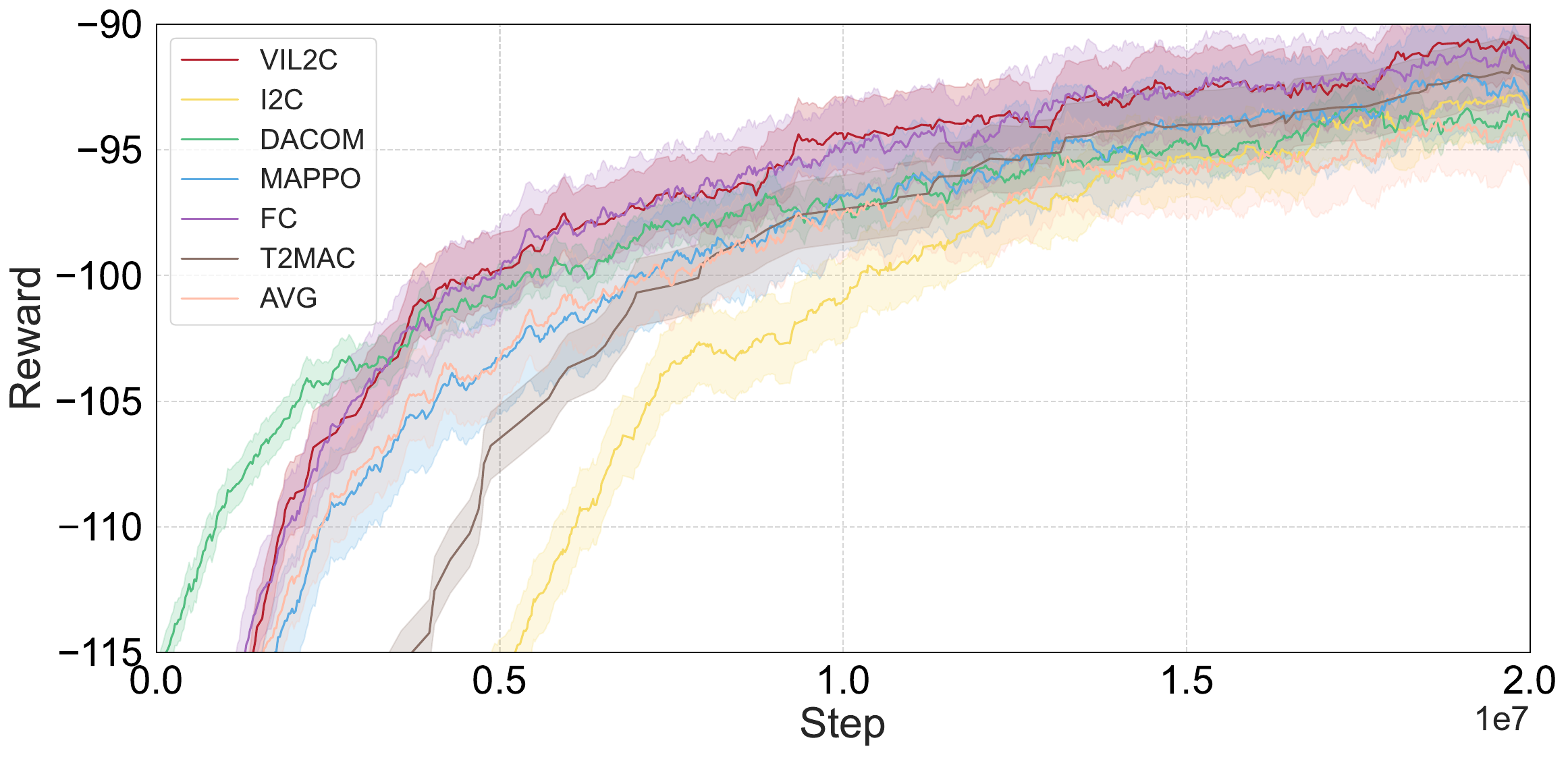}
    \caption{Learning curves for Predator Prey (PP)}
    \label{fig:pp}
\end{figure}

\section{Experiments}
We compare the performance of VIL2C with baselines on three well-known partially-observed multi-agent cooperative  tasks: PP, CN\cite{lowe:mpe}, and SMACv2 \cite{ellis:smacv2}. We also demonstrate the performance of VIL2C under various communication conditions. In addition, as a crucial hyperparameter in VIL2C, the effect of maximum waiting time on performance is explored. 

The baselines include MAPPO, I2C \cite{ding:i2c}, DACOM \cite{yuan:dacom},  T2MAC \cite{sun:t2m2c}, latency-free full communication (FC), and
equal resource allocation with fixed communication waiting (AVG). MAPPO is the CTDE backbone without inter-agent communication. I2C is the individual communication and each agent has a fixed waiting duration. DACOM dynamically adjusts the waiting duration based on the network status. In T2MAC, agents craft personalized messages, select communication windows and partners. In FC, each agent receives all agents' messages without latency. In AVG, agents equally allocate resources and the waiting duration is fixed. We set the fixed waiting time as $30\%$ of total time in baselines.


To incorporate the influence of communication latency on action execution, we modify the original environment by dividing the execution time into two parts: the first part is the time for waiting messages and the second part is the time for action execution. In the waiting time, the agent performs the action selected at last time step. In the execution time, the agent selects and executes a new action.

\subsection{Performance in Different Environments}

\textbf{Predator Prey.} \: In PP experiments, we consider that four slower predators chase two faster preys and two landmarks obstruct their paths. Each predator has an unique observation range, in which it observes landmarks' positions, the positions and velocities relative to other predators and preys. The reward is the sum of all predators’ negative distances to their closest preys or landmarks. Predators are penalized for colliding with others. In this environment, we compare VIL2C with baselines, as shown in Fig.~\ref{fig:pp}. It is seen that the average reward of VIL2C is close to FC that is latency-free and outperforms other baselines. It implies that VIL2C can effectively alleviate the impact of communication latency on cooperation performance. Moreover, it is observed that the rewards of I2C and AVG are lower than that of MAPPO. It indicates that without a judiciously designed low-latency communication system in MARL, the impact of communication latency may overwhelm benefits of communication. 
\begin{figure}[t]
    \centering
    \includegraphics[width=0.95\linewidth]{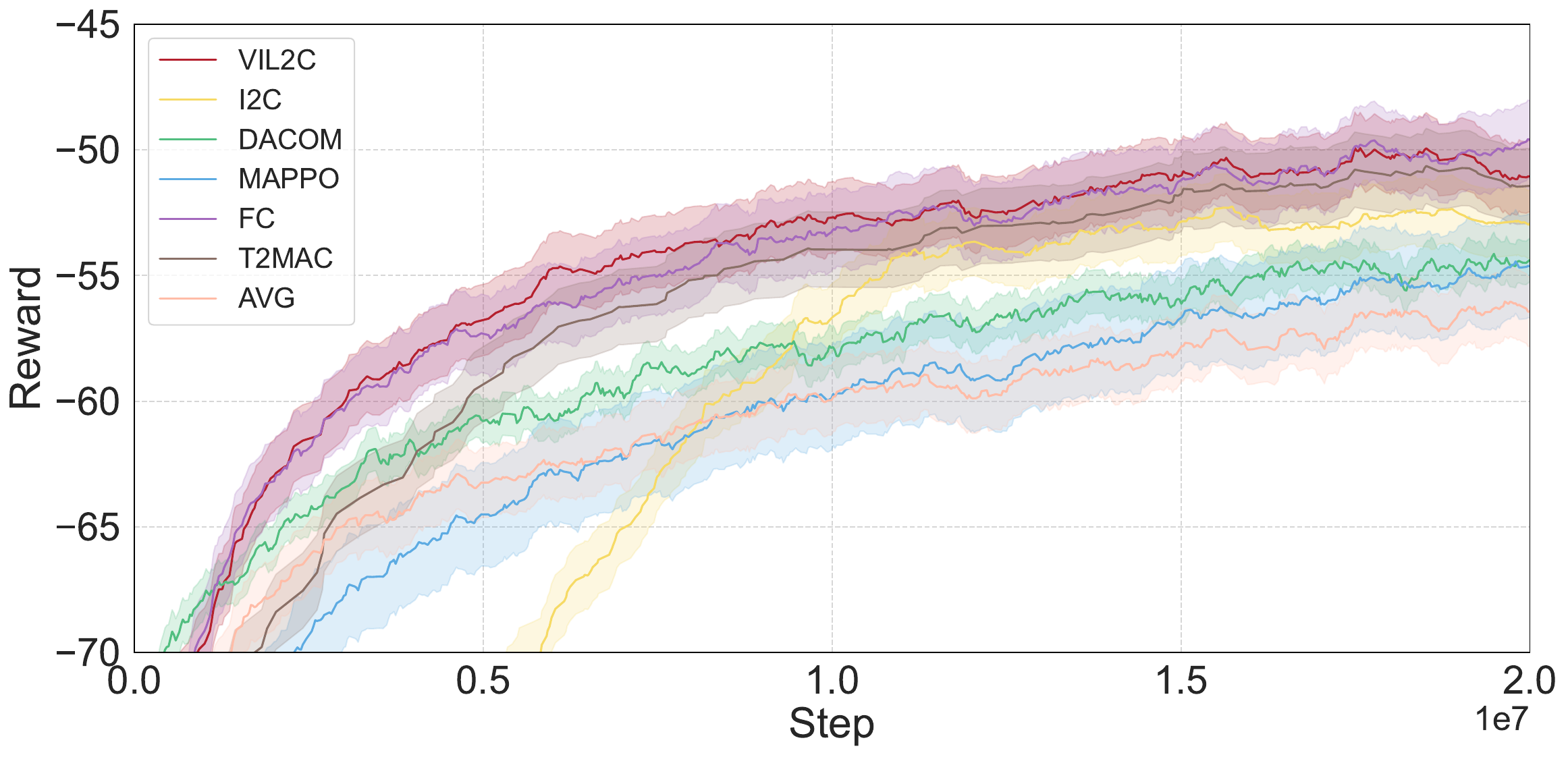}
    \caption{Learning curves for Cooperative Navigation (CN)}
    \label{fig:cn}
\end{figure}
\begin{figure}[b]
    \centering
    \includegraphics[width=0.95\linewidth]{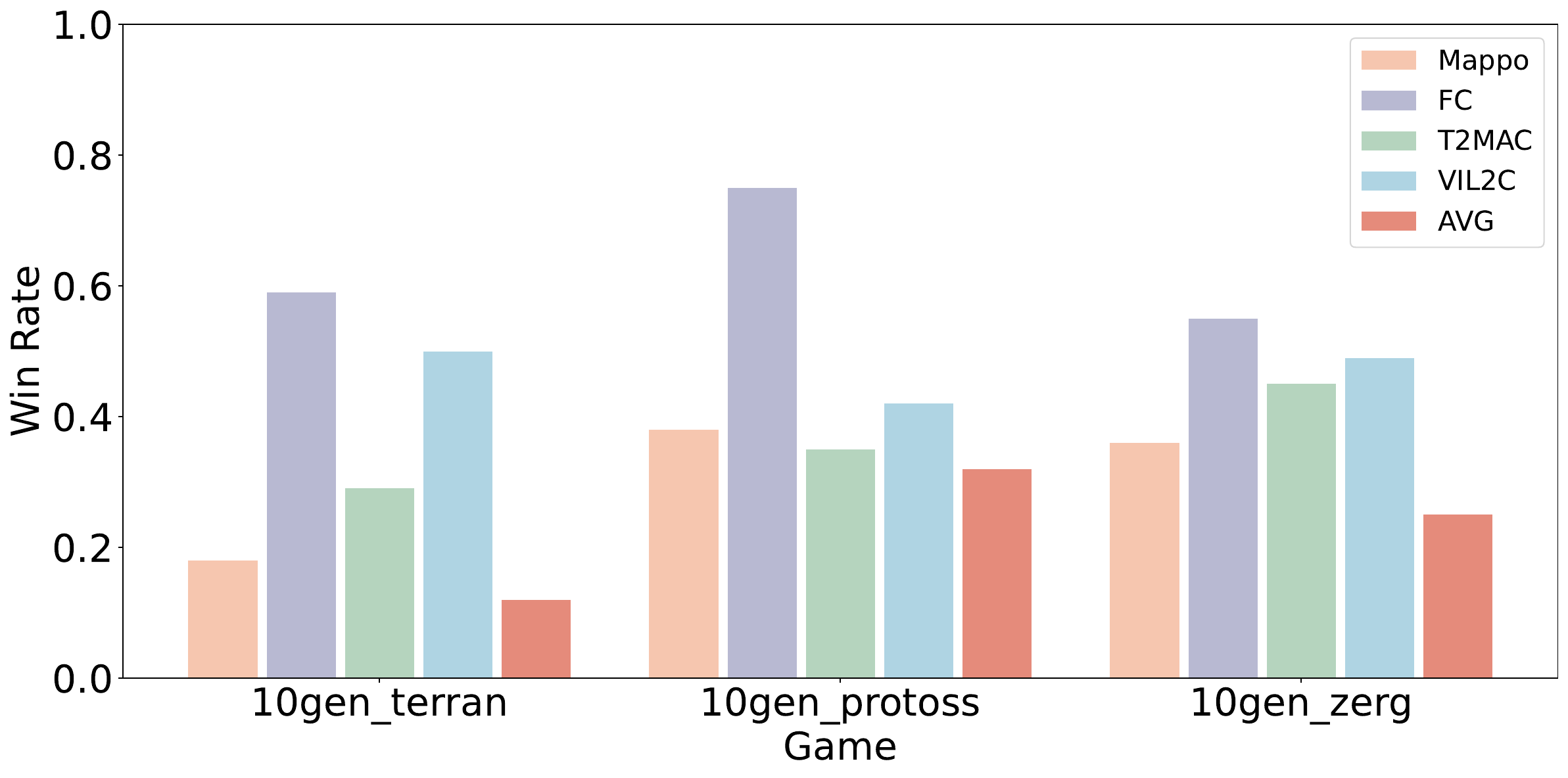}
    \caption{Win Rates for SMACv2}
    \label{fig:smacv2}
\end{figure}
\begin{figure*}[t]
    \centering
    \includegraphics[width=1\linewidth]{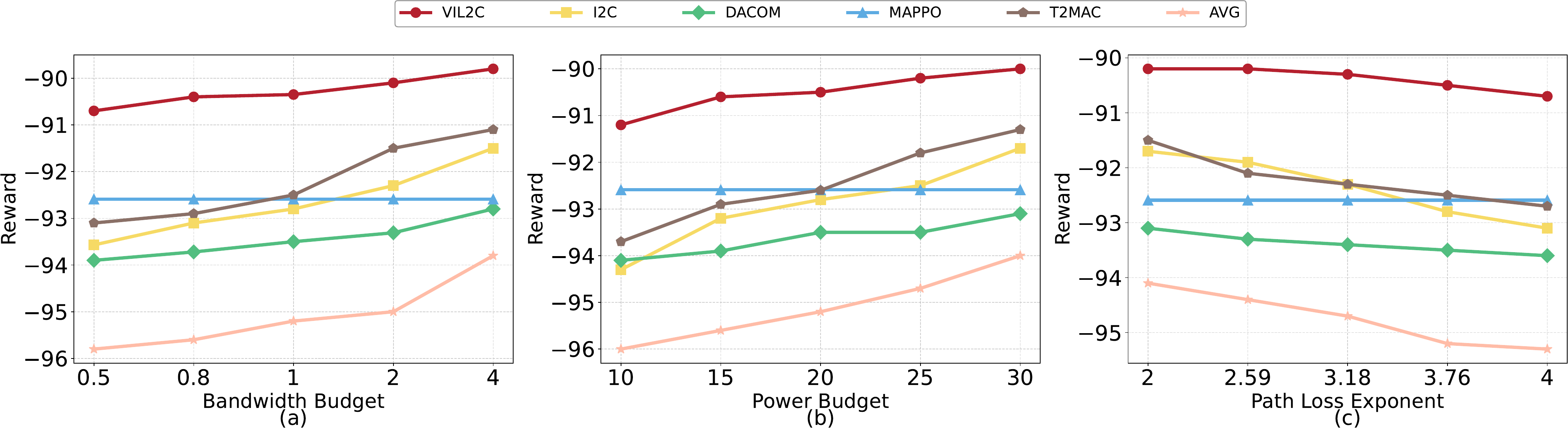}
    \caption{Performance comparisons under different communication conditions.}
    \label{fig:three}
\end{figure*}

\noindent \textbf{Cooperative Navigation.} \: In CN experiments, we consider that four agents aim to occupy four stationary landmarks individually, where the positions of landmarks and agents are randomly initialized. Each agent has an uniquely limited observation range. 
Due to the limitation of observations, agents need to cooperatively reschedule their targeted landmarks, when more than one agent choose the same landmark as the target. We compare the training performance of VIL2C with baselines in this environment, as shown in Fig.~\ref{fig:cn}. It is seen that VIL2C approaches FC that is latency-free and outperforms other baselines.

\noindent \textbf{SMACv2.} \:In SMACv2 experiments, we consider that five agents controlled by the learned algorithm aim to cooperatively attack five enemies via some strategies. Three different combat scenarios are employed, i.e., 10gen\_terran, 10gen\_protoss, and 10gen\_zerg. We compare VIL2C with baselines in the three scenarios, in terms of the win rate, as shown in Fig.~\ref{fig:smacv2}. Here, the win rate is averaged over $32$ test games. It can be seen that FC achieves the highest win rate, due to the reception of messages from all other agents with no communication latency. Moreover, VIL2C outperforms other baselines. This is owing to the communication latency distribution adjustment in VIL2C. Note that, we transform continuous latency into discrete game steps in this environment, where eight game steps equal one MARL step.

\begin{figure}[b]
    \centering
    \includegraphics[width=0.95\linewidth]{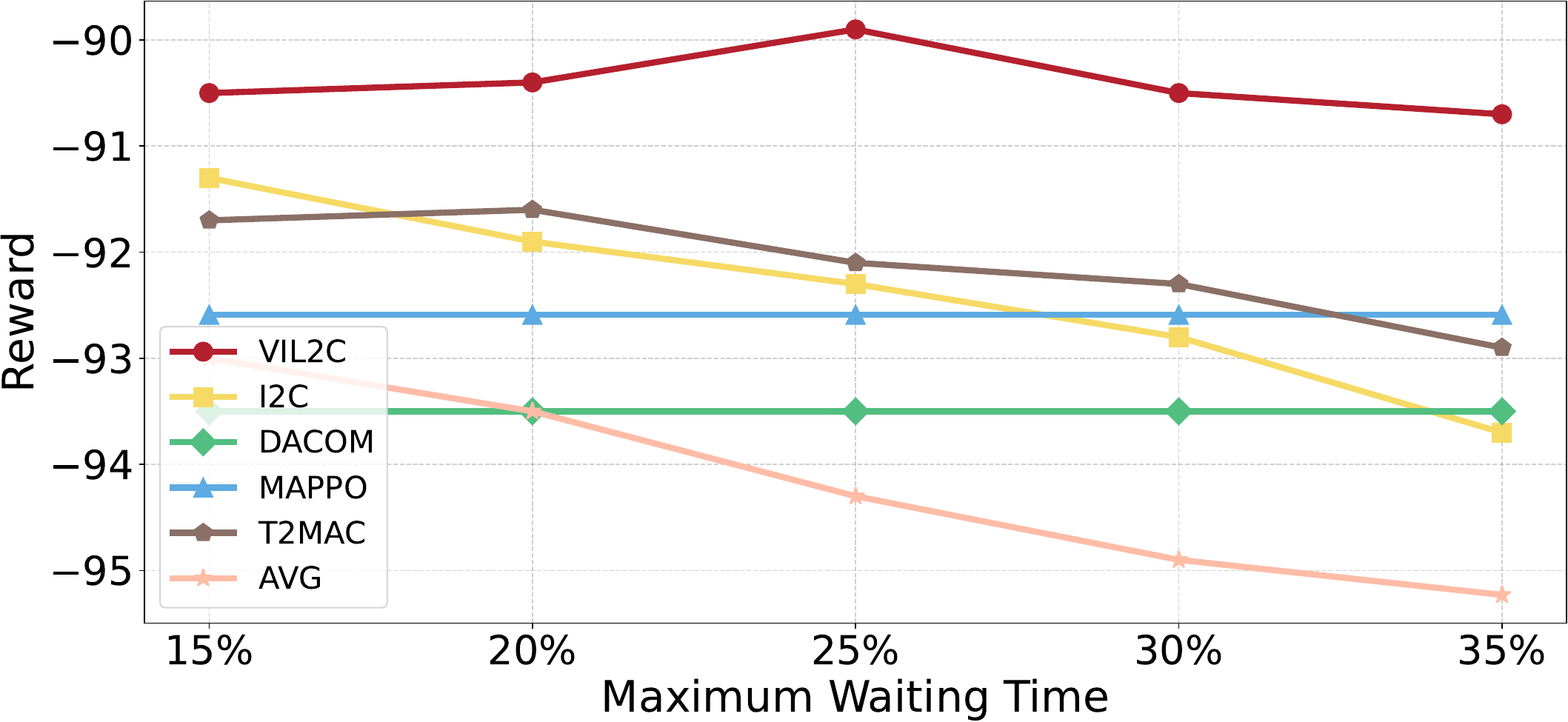}
    \caption{The effect of maximum waiting time.}
    \label{fig:fd}
\end{figure}
\subsection{Performance under Different Communication Conditions}

We evaluate the performance of VIL2C under different communication conditions, as shown in Fig. \ref{fig:three}. Here, we employ the PP task as an example and similar results can be obtained in other tasks. The considered communication conditions include each agent's bandwidth budget, power budget, and the path loss exponent. It can be seen from Figs. \ref{fig:three}(a) and \ref{fig:three}(b) that VIL2C achieves a significant improvement over baselines for the whole range of bandwidth/power budget. Moreover, as the bandwidth/power budget decreases, the performance of VIL2C suffers the slighter degradation than baselines, especially I2C and T2MAC. It implies that VIL2C exhibits great robustness against limited resource budgets. This is because that VIL2C can adaptively adjust the bandwidth/power allocation and the waiting duration for different communication latency. It is also observed from Fig. \ref{fig:three}(c) that VIL2C outperforms baselines during the range of path loss exponent. As the path loss exponent increases, i.e., the channel condition becomes worse, VIL2C illustrates the small performance degradation.

\subsection{Effect of Maximum Waiting Time}

We further explore the effect of maximum waiting time, as shown in Fig. \ref{fig:fd}. As a crucial hyperparameter in MARL with communication latency, the maximum waiting time strikes the balance between the benefit of communication and the cost of message waiting. It can be seen that the performance of VIL2C first rises and then drops, as the maximum waiting time increases. This is because that the increase of waiting time allows agents to receive more messages to improve the performance, while the too large waiting time makes the impact of communication latency overwhelm benefits of communication. 
\begin{table}[H]
\setlength{\tabcolsep}{1.5mm}
\centering
\begin{tabular}{lccc}
\toprule
\textbf{Maps} & \textbf{VIL2C} & \textbf{VIL2C w/o P} & \textbf{VIL2C w/o R} \\ 
\midrule
10gen\_terran & 0.55 & 0.35 & 0.28 \\
10gen\_protoss & 0.42 & 0.32 & 0.26 \\
10gen\_zerg & 0.48 & 0.28 & 0.41 \\
\bottomrule
\end{tabular}
\caption{Ablation for ResoNet and Progressive Reception.}
\label{tab:ablation}
\end{table}
\subsection{Ablation}
To quantify the individual contributions of ResoNet (R) and Progressive Reception (P), we conduct the ablation study in three combat scenarios of SMACv2 environment, as reported in Tab.\ref{tab:ablation}. It is observed that removing either module of VIL2C reduces the win rate, confirming that both are essential. In the scenarios of 10gen\_terran and 10gen\_zerg, the removal of ResoNet demonstrates more severe performance degradation than the removal of progressive reception. Because in these two environments, the agents have smaller observation ranges and the communication gain, especially from important messages, is more obvious. In this case, using resource allocation can reduce the latency of important messages, thereby significantly improving multi-agent cooperation performance. 

\section{Conclusion}
This work investigates proactive communication latency distribution adjustment to mitigate the latency impact on cooperation performance in MARL. To achieve this, we propose VIL2C. Transmitter agents optimize bandwidth and power to maximize the total VoI, while recipient agents adaptively determine the waiting time based on whether sufficient information is obtained. We demonstrate the advantage of VIL2C through theoretical analysis and experiments. However, our focus is on point-to-point communications, where latency is primarily determined by wireless signal propagation, and the latency caused by network conditions, such as network congestion, is not considered in this work. Moreover, a more generalized VoI definition for tasks with various latency sensitivities will be considered for future work.


\section{Acknowledgments}
This work was supported by the National Key R\&D Program of China (No. 2024YFB4505502), the National Natural Science Foundation of China (No. 62532009, No. 62302396), the Hong Kong Research Grants Council under the Areas of Excellence Scheme Grant AoE/E-601/22-R and the NSFC/RGC Collaborative Research Scheme Grant CRS\_HKUST603/22.

\bibliography{aaai2026}

\end{document}